\documentclass{article}
\usepackage{graphicx}
\usepackage{cite}
\usepackage{rotating}

\begin{document}

\title{Information Integration from Distributed Threshold-Based Interactions}

\author{Valmir~C.~Barbosa\thanks{valmir@cos.ufrj.br.}\\
\\
Programa de Engenharia de Sistemas e Computa\c c\~ao, COPPE\\
Universidade Federal do Rio de Janeiro\\
Caixa Postal 68511\\
21941-972 Rio de Janeiro - RJ, Brazil}

\date{}

\maketitle

\begin{abstract}
We consider a collection of distributed units that interact with one another
through the sending of messages. Each message carries a positive ($+1$) or
negative ($-1$) tag and causes the receiving unit to send out messages as a
function of the tags it has received and a threshold. This simple model
abstracts some of the essential characteristics of several systems used in the
field of artificial intelligence, and also of biological systems epitomized by
the brain. We study the integration of information inside a temporal window as
the model's dynamics unfolds. We quantify information integration by the total
correlation, relative to the window's duration ($w$), of a set of random
variables valued as a function of message arrival. Total correlation refers to
the rise of information gain above and beyond that which the units already
achieve individually, being therefore related to consciousness studies in some
models. We report on extensive computational experiments that explore the
interrelations of the model's parameters (two probabilities and the threshold),
highlighting relevant scenarios of message traffic and how they impact the
behavior of total correlation as a function of $w$. We find that total
correlation can occur at significant fractions of the maximum possible value and
provide semi-analytical results on the message-traffic characteristics
associated with values of $w$ for which it peaks. We then reinterpret the
model's parameters in terms of the current best estimates of some quantities
pertaining to cortical structure and dynamics. We find the resulting
possibilities for best values of $w$ to be well aligned with the time frames
within which percepts are thought to be processed and eventually rendered
conscious.

\end{abstract}

\newpage
\section{Introduction}
\label{intro}

A threshold-based system is a collection of loosely coupled units, each
characterized by a state function that depends on how the various inputs to the
unit relate to a threshold parameter. The coupling in question refers to how
the units interrelate, which is by each unit communicating its state to some of
the other units whenever that state changes. Given a set of timing assumptions
and how they relate to the exchange of states among units as well as to state
updates, each individual unit processes inputs (states communicated to it by
other units) and produces a threshold-dependent output (its own new state, which
gets communicated to other units).

The quintessential threshold-based system is undoubtedly the brain, where each
neuron's state function determines whether an action potential is to be fired
down the neuron's axon. This depends on how the combined action potentials the
neuron perceives through the synapses connecting other neurons' axons to its
dendrites (its synaptic potentials) relate to its threshold potential
\cite{bcp15}. The greatly simplified model of the natural neuron known as the
McCulloch-Pitts neuron \cite{mp43}, introduced over seventy years ago, retained
this fundamental property of being threshold-based, and so did generalizations
thereof such as generalized Petri nets \cite{prc81} and threshold automata
\cite{i01}. In fact, this holds for much of the descent from the McCulloch-Pitts
neuron, which has extended through the present day in a succession of ever more
influential dynamical systems.

Such descent includes the essentially deterministic Hopfield networks of the
1980s \cite{h82,h84} and moves on through generalizations of those networks'
Ising-model type of energy function and the associated need for stochastic
sampling. The resulting networks include the so-called Boltzmann machines
\cite{ahs85} and Bayesian networks \cite{p88,h90}, as well as the more general
Markov (or Gibbs) Random Fields \cite{ks80,gg84,b93} and several of the
probabilistic graphical models based on them \cite{kf09}. A measure of the
eventual success of such networks can be gained by considering, for example, the
restricted form of Boltzmann machines \cite{s86,fi14} used in the construction
of deep belief networks \cite{h09}, as well as some of the other deep networks
that have led to landmark successes in the field of artificial intelligence
recently \cite{mnihetal15,lbh15,silveretal16}.

Our interest in this paper is the study of how information gets integrated as
the dynamics of a threshold-based system is played out. The meaning we attach to
the term information integration is similar to the one we used previously in
other contexts \cite{nb11,cb15}. Given a certain amount of time $w$ and a set of
random variables, each related to the firing activity of each of the system's
units inside a temporal window of duration $w$, we quantify integrated
information as the amount of information the system generates as a whole
(relative to a global state of maximum entropy) beyond that which accounts for
the aggregated information the units generate individually (now relative to
local states of maximum entropy). This surplus of information is known as total
correlation \cite{w60} and is fundamentally dependent on how the units interact
with one another.

Our understanding of information integration, therefore, lies in between those
used by approaches that seek it in the synchronization of input/output signals
(see \cite{zmtx15}, for example) and those that share our view but would
consider not just the whole and the individual units but all partitions in
between as well \cite{bt08}. By virtue of this, we remain aligned with the
latter theory by acknowledging that information gets integrated only when it is
generated by the whole in excess of the total its parts can generate
individually. On the other hand, by sticking with total correlation as an
information-theoretic quantity that requires only two partitions of the set of
units to be considered (one that is fully cohesive and another that is maximally
disjointed), we ensure tractability way beyond that of the all-partitions
theory.

We conduct all our study on a simple model of threshold-based systems. In this
model, the units are placed inside a cube and exchange messages whose delivery
depends on their propagation speed and the Euclidean distance between sender and
receiver. Every message is tagged, and upon reaching its destination, its tag
is used to move a local accumulator either toward or away from the threshold.
Reaching the threshold makes the unit send out messages and the accumulator is
reset. There are three parameters in the model. Two of them are probabilities
(that a message is tagged so that the accumulator at the destination gets
decreased upon its arrival, and that a unit sends a message to each of the other
units), the other being the value of the threshold. Parameter values are the
same for all units.

This model is by no means offered as an accurate representation of any
particular threshold-based system, but nevertheless summarizes some key aspects
of such systems through its relatively few parameters. In particular, it gives
rise to three possible expected global regimes of message traffic. One of them
is perfectly balanced, in the sense that on average as much traffic reaches the
units as leaves them. In this case, message traffic is sustained indefinitely.
In each of the other two regimes, by contrast, either more traffic reaches the
units than leaves them, or conversely. Message traffic dies out in the former of
these two (unless the units receive further external input) but grows
indefinitely in the latter one.

We find that information integration is guaranteed to occur at high levels for
some window durations whenever message traffic is sustained at the
perfect-balance level or grows. We also find that this happens nearly
independently of parameter variations. On the other hand, we also find that
information integration is strongly dependent on the model's parameters, with
significant levels occurring only for some combinations, whenever message
traffic is imbalanced toward the side that prevents it from being sustained.
Here we once again turn to the brain, whose cortical activity is in many
accounts characterized as tending to be sparse \cite{kgh05,sos06}, as an
emblematic example.

We proceed as follows. Our message-passing model is laid out in
Section~\ref{model}, where its geometry and distributed algorithm are detailed
and the question of message imbalance is introduced. An account of our use of
total correlation is given in Section~\ref{tc}, followed by our methodology in
Section~\ref{methods}. Results, discussion, and conclusion follow, respectively
in Sections~\ref{results}, \ref{disc}, and~\ref{concl}.

\section{Model}
\label{model}

Our system model comprises a structural component and an algorithmic one. The
two are described in what follows, along with some analysis of how they
interrelate.

\subsection{Underlying geometry}
\label{geom}

For $1\le D\le 3$, our model is based on $N$ simple processing units, henceforth
referred to as nodes, each one placed at a fixed position inside the
$D$-dimensional cube of side $\ell$. The position of node $i$ has coordinates
$X_i^{(1)},\ldots,X_i^{(D)}$, so the Euclidean distance between nodes $i$ and
$j$ is
\begin{equation}
\delta_{ij}=\sqrt{(X_i^{(1)}-X_j^{(1)})^2+\cdots+(X_i^{(D)}-X_j^{(D)})^2}.
\end{equation}
We assume that nodes can communicate with one another by sending messages that
propagate at the fixed speed $\sigma$ on a straight line. The delay incurred by
a message sent between nodes $i$ and $j$ in either direction is therefore
$\delta_{ij}/\sigma$.

Our computational experiments will all be such that the $N$ nodes are placed in
the $D$-dimensional cube uniformly at random. In this case, and in the limit of
infinite $N$, the expected distance between two randomly chosen nodes $i$ and
$j$ is given by
\begin{equation}
\Delta_\ell^{(D)}=
\left(\frac{1}{\ell}\right)^{2D}
\int_0^\ell\cdots\int_0^\ell\delta_{ij}
\mathrm{d}X_i^{(1)}\cdots\mathrm{d}X_i^{(D)}
\mathrm{d}X_j^{(1)}\cdots\mathrm{d}X_j^{(D)},
\end{equation}
where $1/\ell$ is the probability density for each of the $2D$ variables.
Letting $X_k^{(d)}=\ell x_k^{(d)}$ in this equation for $k\in\{i,j\}$ and
$d\in\{1,\ldots,D\}$ yields
\begin{eqnarray}
\Delta_\ell^{(D)}=
\frac{\ell^{2D+1}}{\ell^{2D}}
\int_0^1\cdots\int_0^1\delta_{ij}
\mathrm{d}x_i^{(1)}\cdots\mathrm{d}x_i^{(D)}
\mathrm{d}x_j^{(1)}\cdots\mathrm{d}x_j^{(D)},
\end{eqnarray}
where $\delta_{ij}$ now has $x$'s in place of the $X$'s. We then have
\begin{equation}
\Delta_\ell^{(D)}=\ell\Delta_1^{(D)},
\end{equation}
with the expected distances in the unit cube for the numbers of dimensions of
interest being well known: $\Delta_1^{(1)}=1/3$, $\Delta_1^{(2)}\approx 0.5214$
\cite{mmp99}, and $\Delta_1^{(3)}\approx 0.6617$ \cite{rb78}.

In addition to expected distances, the associated variances will also at one
point be useful. Analytical expressions for most of them seem to have remained
unknown thus far, but the underlying probability densities have been found to be
more concentrated around the expected values given above as $D$ grows
\cite{z03}. That is, variance is greatest for $D=1$.

\subsection{Network algorithmics}
\label{algo}

We view the $N$ nodes as running an asynchronous message-passing algorithm
collectively. By asynchronous we mean that each node remains idle until a
message arrives. When this happens, the arriving message is processed by the
node, which may result in messages being sent out as well. Such a purely
reactive stance on the part of the nodes requires at least one node to send out
at least one message without having received one, for startup. We assume that
this is done by all nodes initially, after which they start behaving reactively.

We assume that each message carries a signed-unit tag (i.e., either $+1$ or
$-1$) with it. The specific tag to go with the message is chosen
probabilistically by its sender at send time, with $-1$ being chosen with
probability $p_-$. The processing done by node $i$ upon arrival of a message is
the heart of the system's thresholding nature and involves manipulating an
accumulator $A_i$, initially equal to $0$, to which every tag received is added
(unless $A_i=0$ and the tag is $-1$, in which case $A_i$ remains unchanged).
Whenever $A_i$ reaches a preestablished integer value $\tau>0$, node $i$ sends
out messages of its own and $A_i$ is reset to $0$. Thus, the integer $\tau$ acts
as a threshold governing the sending of messages by (the firing of) node $i$.
The values of $p_-$ and $\tau$ are the same for all nodes.

It follows from this simple rule that the value of $A_i$ is perpetually confined
to the interval $[0,\tau]$. The expected number of messages that node $i$ has to
receive in order for $A_i$ to be increased all the way from $0$ to $\tau$ is the
node's expected number of message arrivals between firings, henceforth denoted
by $\mu=\mu(p_-,\tau)$. The value of $\mu$ can be calculated easily once we
recognize that $\mu$ is simply the expected number of steps for the following
discrete-time Markov chain to reach state $\tau$ having started at state $0$.
The chain has states $0,\ldots,\tau$ and transition probability $p_{rs}$, from
state $r$ to state $s$, given by
\begin{equation}
p_{rs}=
\left\{
\begin{array}{ll}
1-p_- &\mbox{if $s=r+1$;}\\
p_- &\mbox{if $s+1=r<\tau$ or $r=s=0$;}\\
1 &\mbox{if $r=s=\tau$;}\\
0 &\mbox{otherwise.}
\end{array}
\right.
\end{equation}
It is easily solved and yields
\begin{equation}
\mu=
\frac{p_-}{(1-2p_-)^2}
\left[\left(\frac{p_-}{1-p_-}\right)^\tau-1\right]+
\frac{\tau}{1-2p_-}
\label{mu}
\end{equation}
for $p_-\neq 0.5$ \cite[page 348]{f68}.\footnote{For $p_-=0.5$ we have
$\mu=\tau(\tau+1)$ \cite[page 349]{f68}, but this holds for none of our
computational experiments.}

The sending of messages when a node fires is based on another parameter,
$p_\to$, which is the probability with which the node sends a message to each of
the other $N-1$ nodes. It follows that the expected number of messages that get
sent out is $(N-1)p_\to$. The value of $p_\to$ is the same for all nodes as
well.

\subsection{Local imbalance and global message traffic}
\label{imb}

At node $i$, a balance exists between message output and message input when the
expected number of messages sent out at each firing is the same as the expected
number of messages received between two successive firings. That is, message
traffic is locally balanced when $(N-1)p_\to=\mu$. It is locally imbalanced
otherwise, which can be quantified by the difference $I$, defined to be
\begin{equation}
I=\frac{(N-1)p_\to}{\mu}-1.
\label{I}
\end{equation}

Given $I$, clearly the instantaneous density of global message output at time
$t$, denoted by $M(t)$, is expected to remain constant with $t$ if $I=0$, or to
decrease or increase exponentially with $t$ depending on whether $I<0$ or $I>0$,
respectively. This behavior is described by
\begin{equation}
\frac{\mathrm{d}M(t)}{\mathrm{d}t}=\frac{I}{T_0}M(t),
\label{dMdt}
\end{equation}
where the $T_0$ participating in the time constant $T_0/I$ is some fundamental
amount of time related to the system's geometry and kinetics. In
Section~\ref{results}, we provide empirical evidence that $T_0$ is the expected
delay undergone by a message, given by
\begin{equation}
T_0=\frac{\Delta_\ell^{(D)}}{\sigma}.
\label{t0}
\end{equation}
Equation~(\ref{dMdt}) is of immediate solution, yielding
\begin{equation}
M(t)=M_0e^{It/T_0},
\end{equation}
where
\begin{equation}
M_0=M(0)=N(N-1)p_\to
\end{equation}
is the expected number of messages that all nodes, collectively, send out
initially.

Similarly, the cumulative global message output inside a temporal window of
duration $w$ starting at time $t$ is
\begin{equation}
M_t(w)=
\int_t^{t+w}M(t)\mathrm{d}t=
\frac{M_0T_0}{I}
\left(e^{I(t+w)/T_0}-e^{It/T_0}\right).
\label{Mtw}
\end{equation}
In the case of locally balanced message traffic ($I=0$), this expression is
easily seen to yield
\begin{equation}
M_t(w)=M_0w,
\end{equation}
therefore independent of $t$. Otherwise, $M_t(w)$ either decreases or increases
exponentially with $t$, depending respectively on whether $I<0$ or $I>0$.

\subsection{Graph-theoretic interpretations}
\label{rgraph}

The model described so far in Section~\ref{model} can be regarded as a directed
geometric graph, that is, a graph whose nodes are positioned in some region of
interest (the $D$-dimensional cube of side $\ell$) and whose edges are directed.
It is moreover a complete graph without self-loops, in the sense that an edge
exists directed from every node $i$ to every node $j\neq i$.

Our use of the model in the sequel will require the nodes to be positioned at
random before each new run of the distributed algorithm of Section~\ref{algo},
so an alternative interpretation that one might wish to consider views the model
as a variation of the traditional random geometric graph \cite{p03}. In this
variation, an edge exists directed from $i$ to $j\neq i$ with fixed probability
$p_\to$, independently of any other node pair. That is, aside from node
positioning the graph underlying our model is an Erd\H{o}s-R\'{e}nyi random
graph \cite{er59} as extended to the directed case \cite{k90}.

This interpretation is somewhat loose, though, because it requires that we view
each individual run of the algorithm as being equivalent to several runs on
independent instances of the underlying random graph, with multiple further runs
serving to validate any statistics that one may come up with at the end. This is
hard to justify, however, particularly when one considers the nonlinearities
characterizing the quantities we will average over all runs of the algorithm
(see Section \ref{tc}). Even so, interpreting our model in terms of random
graphs remains tantalizing in some contexts. For example, it allows the
parameter $p_-$ to be regarded as the fraction of a node's in-neighbors from
which messages with negative tags are received. In the context of networks of
the brain at the neuronal level, for example, an abstract rendering of the
fraction of neurons that are inhibitory is obtained (see Section~\ref{brain}).

\section{Total correlation}
\label{tc}

We use the total correlation of $N$ random variables \cite{w60}, each
corresponding to one of the $N$ nodes, as a measure of information integration.
Each of these variables is relative to a temporal window of fixed duration $w$,
the variable corresponding to node $i$ being denoted by $X_i^{(w)}$ and taking
up values from the set $\{0,1\}$. The intended semantics is that $X_i^{(w)}=1$
if and only if node $i$ receives at least one message in a time interval of
duration $w$. We also use the shorthands $\mathbf{X}^{(w)}$ and $\mathbf{x}$ to
denote the sequence of variables $(X_1^{(w)},\ldots,X_N^{(w)})$ and the sequence
of values $(x_1,\ldots,x_N)$, respectively. Thus, $\mathbf{X}^{(w)}=\mathbf{x}$
stands for the joint valuation $X_1^{(w)}=x_1,\ldots,X_N^{(w)}=x_N$.

Given the marginal Shannon entropy of each variable $X_i^{(w)}$,
\begin{equation}
H(X_i^{(w)})=
-\sum_{x\in\{0,1\}}
\mathrm{Pr}(X_i^{(w)}=x)\log_2\mathrm{Pr}(X_i^{(w)}=x),
\end{equation}
and the joint Shannon entropy
\begin{equation}
H(\mathbf{X}^{(w)})=
-\sum_{\mathbf{x}\in\{0,1\}^N}
\mathrm{Pr}(\mathbf{X}^{(w)}=\mathbf{x})
\log_2\mathrm{Pr}(\mathbf{X}^{(w)}=\mathbf{x}),
\end{equation}
the total correlation of the $N$ variables given $w$ is defined as\footnote{The
$N=2$ case of this formula coincides with that for mutual information, but one
is to note that in the general case the two formulas are completely different
\cite{h80}.}
\begin{equation}
C(\mathbf{X}^{(w)})=
\sum_{i=1}^NH(X_i^{(w)})-H(\mathbf{X}^{(w)}).
\end{equation}

To see the significance of this definition in our context, consider the flat
joint probability mass function,
$\mathrm{Pr}(\mathbf{X}^{(w)}=\mathbf{x})=2^{-N}$ for all
$\mathbf{x}\in\{0,1\}^N$. This mass function entails maximum uncertainty of the
variables' values, hence the maximum possible value of the joint entropy,
$\hat{H}(\mathbf{X}^{(w)})=N$. It also implies flat marginals,
$\mathrm{Pr}(X_i^{(w)}=x)=0.5$ for all $x\in\{0,1\}$ and all $i$, and again the
maximum possible value of each marginal entropy, $\hat{H}(X_i^{(w)})=1$. The
difference from the actual joint entropy $H(\mathbf{X}^{(w)})$ to its maximum
reflects a reduction of uncertainty, or an information gain,
\begin{equation}
G(\mathbf{X}^{(w)})=
\hat{H}(\mathbf{X}^{(w)})-H(\mathbf{X}^{(w)})=N-H(\mathbf{X}^{(w)}),
\end{equation}
the same holding for each of the marginals,
\begin{equation}
G(X_i^{(w)})=
\hat{H}(X_i^{(w)})-H(X_i^{(w)})=1-H(X_i^{(w)}).
\end{equation}

Thus, it is possible to rewrite the expression for $C(\mathbf{X}^{(w)})$ in such
a way that
\begin{equation}
G(\mathbf{X}^{(w)})=C(\mathbf{X}^{(w)})+\sum_{i=1}^NG(X_i^{(w)}).
\end{equation}
That is, the total correlation of all $N$ variables is the information gain
that surpasses their combined individual gains. This surplus is zero if and only
if the variables are independent of one another, that is, precisely when
$\mathrm{Pr}(\mathbf{X}^{(w)}=\mathbf{x})=
\prod_{i=1}^N\mathrm{Pr}(X_i^{(w)}=x_i)$
for all $\mathbf{x}\in\{0,1\}^N$, since in this case we have
$H(\mathbf{X}^{(w)})=\sum_{i=1}^NH(X_i^{(w)})$. It is strictly positive
otherwise, with a maximum possible value of $N-1$.

Achieving this maximum requires a joint probability mass function assigning no
mass to any but two members of $\{0,1\}^N$, say $\mathbf{x}$ and $\mathbf{y}$,
and moreover that these two be equiprobable
(i.e., $\mathrm{Pr}(\mathbf{X}^{(w)}=
\mathbf{x})=\mathrm{Pr}(\mathbf{X}^{(w)}=\mathbf{y})=0.5$)
and complementary to each other (i.e., $x_i+y_i=1$ for every $i$). Referring
back to the intended meaning of the $N$ random variables, total correlation is
maximized in those runs of the distributed algorithm of Section~\ref{algo} for
which a partition $(\mathcal{X}_1,\mathcal{X}_2)$ of the set
$\{X_1^{(w)},\ldots,X_N^{(w)}\}$ exists with the following two properties.
First, no matter which particular window of duration $w$ we concentrate on, the
set of nodes that receive at least one message inside the window is either
$\mathcal{X}_1$ or $\mathcal{X}_2$. Second, the first property holds with
$\mathcal{X}_1$ for exactly half such windows.

While these are exceedingly stringent conditions both spatially and temporally,
perhaps implying that values of total correlation equal to or near $N-1$ are
practically unachievable, they serve to delineate those scenarios with a chance
of generating substantial amounts of total correlation. Specifically, such
scenarios will on average have a pattern of global message traffic, inside a
window, that is neither too sparse nor too dense. Furthermore, sustaining such
an amount of total correlation as time elapses will also require traffic
patterns that deviate only negligibly from the ones yielding the average,
possibly entailing some variability on window sizes. Our methodology to track
and validate values of $w$ leading to noteworthy total correlation is described
next. It involves computational experiments for a variety of values for $w$, and
also gauging the cumulative global message output $M_t(w)$ that results from
each experiment against a function of the reference number of messages and
reference delay embodied in $M_0$ and $T_0$, respectively.

\section{Methods}
\label{methods}

Our results are based on running the distributed algorithm of
Section~\ref{algo} for a fixed geometry (i.e., fixed number of dimensions
$D\in\{1,2,3\}$, fixed value of the cube side $\ell$, and fixed positioning of
$N$ nodes in the cube) and a fixed set of values for the parameters ($p_-$,
$\tau$, and $p_\to$). Each run of the algorithm terminates either when no more
message is in transit (so none will ever be thenceforth, given the algorithm's
reactive nature) or when a preestablished maximum number of messages in transit
has been reached, whichever comes first. Imposing the latter upper bound is
important because it serves to size the data structures where messages in
transit are kept for later processing.

Node positioning is achieved uniformly at random, so multiple runs are needed
for each fixed configuration $(D,\ell,N,p_-,\tau,p_\to)$. Each run leaves a
trace of all events taking place as it unfolds, each event referring to the
arrival of a message at a node and comprising the node's identification and the
message's arrival time. A series of values for $w$ is then considered and for
each $w$ each of the traces is analyzed, yielding the total correlation produced
by the corresponding run. The average total correlation over all the runs is
then reported.

Following our discussion at the end of Section~\ref{tc}, for each value of $w$
we gauge Eq.~(\ref{Mtw}) against the approximation to it given by
$M_0(T_0)=M_0T_0$, according to which $M_0$ messages get sent at time $t=0$ and
received at time $T_0$. We do this by postulating a proportionality constant
$\alpha>0$ between them, that is, by assuming 
\begin{equation}
M_t(w)=\alpha M_0T_0.
\label{aM0T0}
\end{equation}
Doing this allows us to express $w$ as a function of $\alpha$ for each $t$ and,
whenever possible, to characterize traffic regimes giving rise to substantial
amounts of total correlation.

\subsection{Supporting analysis}
\label{analysis}

We denote the value of $t$ upon termination of a run by $T$ and the total number
of messages sent by $M$. An approximation to Eq.~(\ref{Mtw}) similar to the one
above can be used to relate $T$ and $M$ as $M_0(T)=MT_0$, whose right-hand side
quantifies what would be expected to happen if all $M$ messages were sent at
time $t=0$ and received at time $T_0$. This leads to
\begin{equation}
T=\frac{T_0}{I}\ln\left(1+\frac{MI}{M_0}\right).
\end{equation}

Solving Eq.~(\ref{aM0T0}) for $w=w(t)$ given $\alpha$ yields
\begin{equation}
w(t)=\frac{T_0}{I}\ln\left(1+\alpha Ie^{-It/T_0}\right),
\label{wt}
\end{equation}
whose value for $t=0$ is the duration of the first window,
\begin{equation}
w_0=w(0)=\frac{T_0}{I}\ln(1+\alpha I).
\label{w0}
\end{equation} 
As for the duration of the last window, which we denote by $w_*$, it can
likewise be found by solving Eq.~(\ref{aM0T0}) for $w$, now letting the window's
start time be $t=T-w$, then letting $w_*=w$. We obtain
\begin{equation}
w_*=\frac{T_0}{I}\ln\left(\frac{1+MI/M_0}{1+(M/M_0-\alpha)I}\right).
\label{w*}
\end{equation}

The average window duration between time $t=0$ and time
\begin{equation}
T-w_*=\frac{T_0}{I}\ln\left(1+\left(\frac{M}{M_0}-\alpha\right)I\right),
\end{equation}
denoted by $\bar{w}$, is also of interest and comes from the indefinite integral
\begin{equation}
f(t)=
\int w(t)\mathrm{d}t=
\left(\frac{T_0}{I}\right)^2\mathrm{Li}_2\left(-\alpha Ie^{-It/T_0}\right),
\end{equation}
where $\mathrm{Li}_2(z)=\sum_{k\ge 1}z^k/k^2$ is the dilogarithm of $z$. This
given, we obtain
\begin{eqnarray}
\bar{w}
&=&\frac{f(T-w_*)-f(0)}{T-w_*}\nonumber\\
&=&\frac{T_0}{I}\;
\frac{\mathrm{Li}_2(-\alpha I/[1+\left(M/M_0-\alpha\right)I])-
\mathrm{Li}_2(-\alpha I)}
{\ln(1+\left(M/M_0-\alpha\right)I)}.
\label{barw}
\end{eqnarray}

While for $I=0$ we have $T=MT_0/M_0$ and
\begin{equation}
w_0=\bar{w}=w_*=\alpha T_0,
\label{aforI=0}
\end{equation}
for $I\neq 0$ everything depends on the sign of $I$. If $I<0$, then we need
$M/M_0<-1/I$ in order for $T$ to be well defined. Moreover, we have
\begin{equation}
0<w_0<\bar{w}<w_*<T,
\label{wforIneg}
\end{equation}
where the first inequality holds if $\alpha<M/M_0$, this being necessary and
sufficient for the last inequality to hold as well. For $I>0$, on the other
hand, $T$ is always well defined and we get
\begin{equation}
T>w_0>\bar{w}>w_*>0.
\label{wforIpos}
\end{equation}
In this case, the constraint $\alpha<M/M_0$ is necessary and sufficient for both
the first and the last inequality to hold.

\subsection{Computational experiments}
\label{exp}

We organize our computational experiments into settings, numbered I--IV, each
comprising all configurations $(D,\ell,N,p_-,\tau,p_\to)$ for which $N$, $p_-$,
and $\tau$ are fixed. In each setting, there are three possibilities for the
value of $p_\to$, one ensuring $I<0$ ($p_\to=0.01$), one for $I=0$ (extracted
from Eq.~(\ref{I})), and one for $I>0$ ($p_\to=0.06$). The four settings are
summarized in Table~\ref{table1}. Each of settings~II--IV is derived from
setting~I by a change in the value of $N$, $p_-$, or $\tau$, respectively.

\begin{table}
\caption{Parameter values for each of settings~I--IV and each of the three
possibilities for the value of $I$ in relation to $0$. Values of $p_\to$ are
given by $p_\to=\mu/(N-1)$ for $I=0$, as per Eq.~(\ref{I}).}
\label{table1}
\centering
\begin{tabular}{ccccccc}
\hline
Setting  &$N$  &$p_-$  &$\tau$  &&$p_\to$\\
\cline{5-7}
&&&&$I<0$  &$I=0$  &$I>0$\\
\hline
I  &500  &0.30  &5  &0.01  &0.02135  &0.06\\
II  &300  &0.30  &5  &0.01  &0.03563  &0.06\\
III  &500  &0.18  &5  &0.01  &0.01478  &0.06\\
IV  &500  &0.30  &10  &0.01  &0.04634  &0.06\\
\hline
\end{tabular}

\end{table}

Each setting entails $2\,100$ runs of the algorithm for each $p_\to$ value, this
total comprising $100$ independent trials for each combination of
$D\in\{1,2,3\}$ and $\ell\in\{10^{-3},10^{-2},\ldots,10^3\}$. Each run starts by
placing the $N$ nodes anew, uniformly at random, and proceeds therefrom subject
to the further indeterminacies that characterize the algorithm through the
probabilities $p_-$ and $p_\to$. Messages are assumed to travel at the speed
$\sigma=1$. Each of the $100$ traces resulting from the same configuration is
then analyzed and the corresponding value of $C(\mathbf{X}^{(w)})$ is computed,
the average over the traces being reported at the end. This is done for each
$w\in\{2^{-33},2^{-32.5},\ldots,2^{10}\}$.

One last important aspect of a run has to do with the determination of the
maximum number of messages ever allowed to be in transit as the run unfolds. We
determine this number via the formula $1\,000N\mu$, where $1\,000\mu$ can be
interpreted as the expected number of messages a node must receive if it is to
fire $1\,000$ times during the run. This latter number is in many ways
arbitrary, though, having to do only with ensuring that the computational
resources required by all runs remain manageable. As a result, taken as a whole
the $25\,200$ runs have generated a total of about 394 gigabytes of trace in
compressed format. Moreover, processing these data for the determination of the
various $C(\mathbf{X}^{(w)})$ averages has required several weeks of computation
on an Intel Xeon E5-1650 core clocked at 3.2 GHz and having exclusive access to
30 gigabytes of RAM. 

\section{Results}
\label{results}

\begin{figure}
\centering
\includegraphics[scale=0.54]{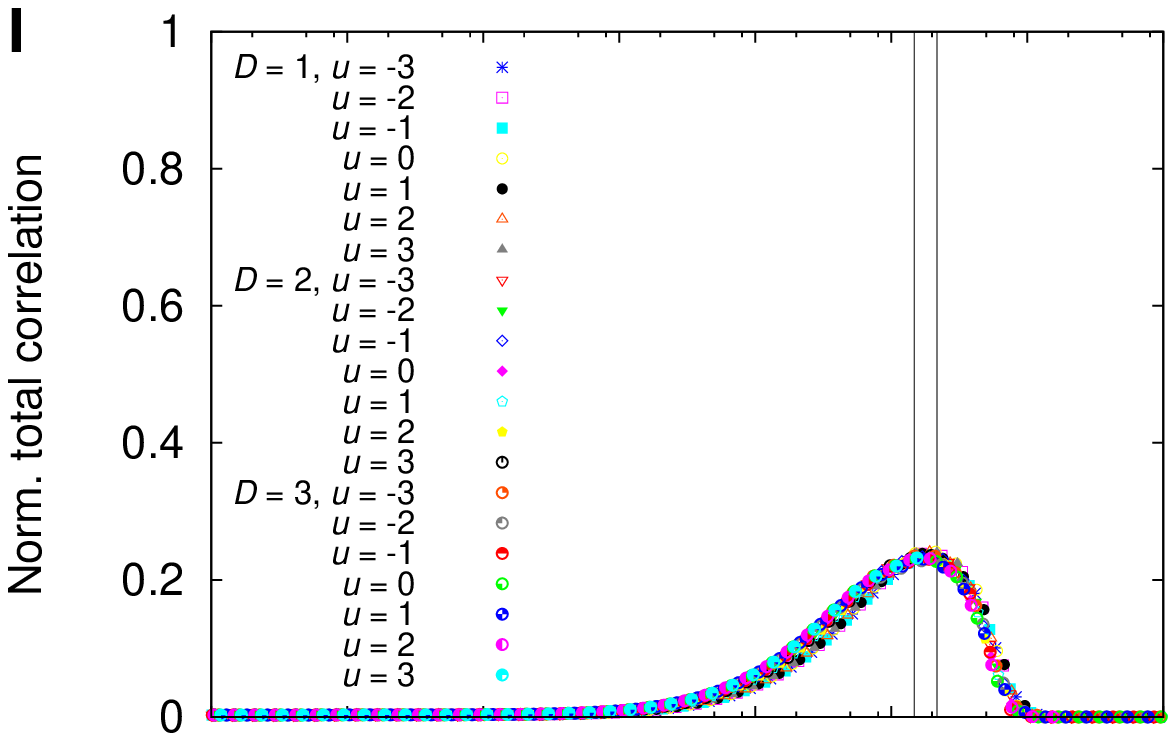}\\
\vspace{-20pt}
\includegraphics[scale=0.54]{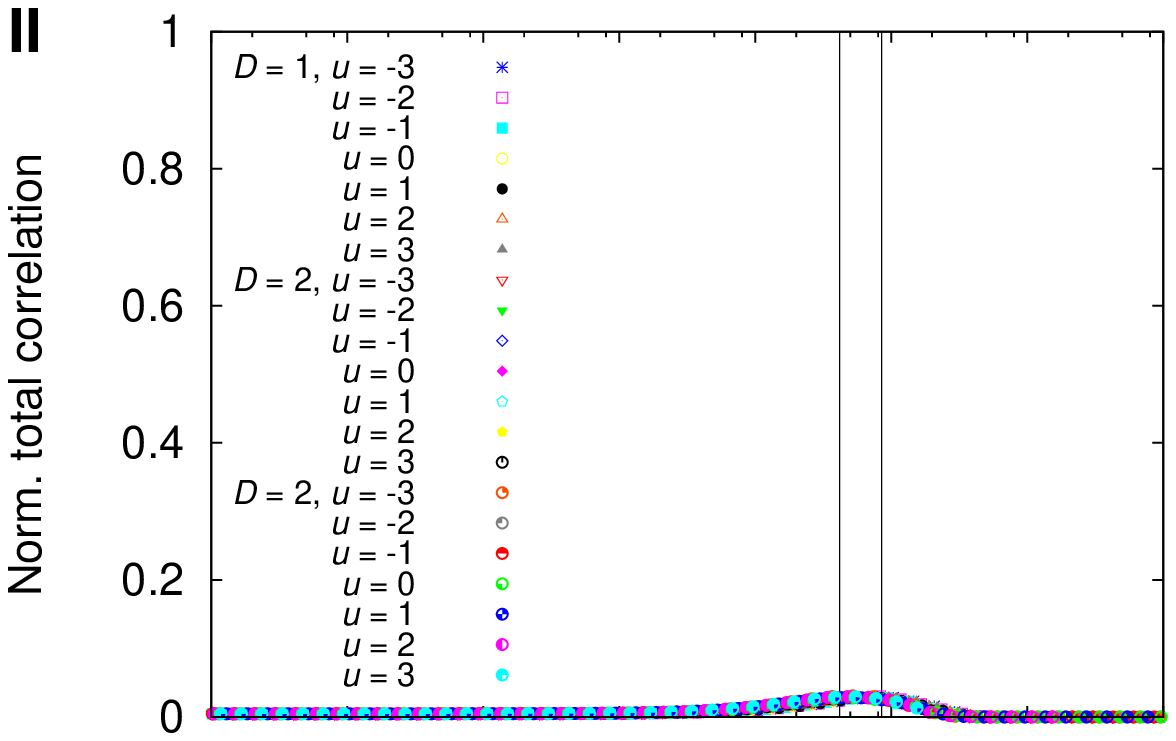}\\
\vspace{-20pt}
\includegraphics[scale=0.54]{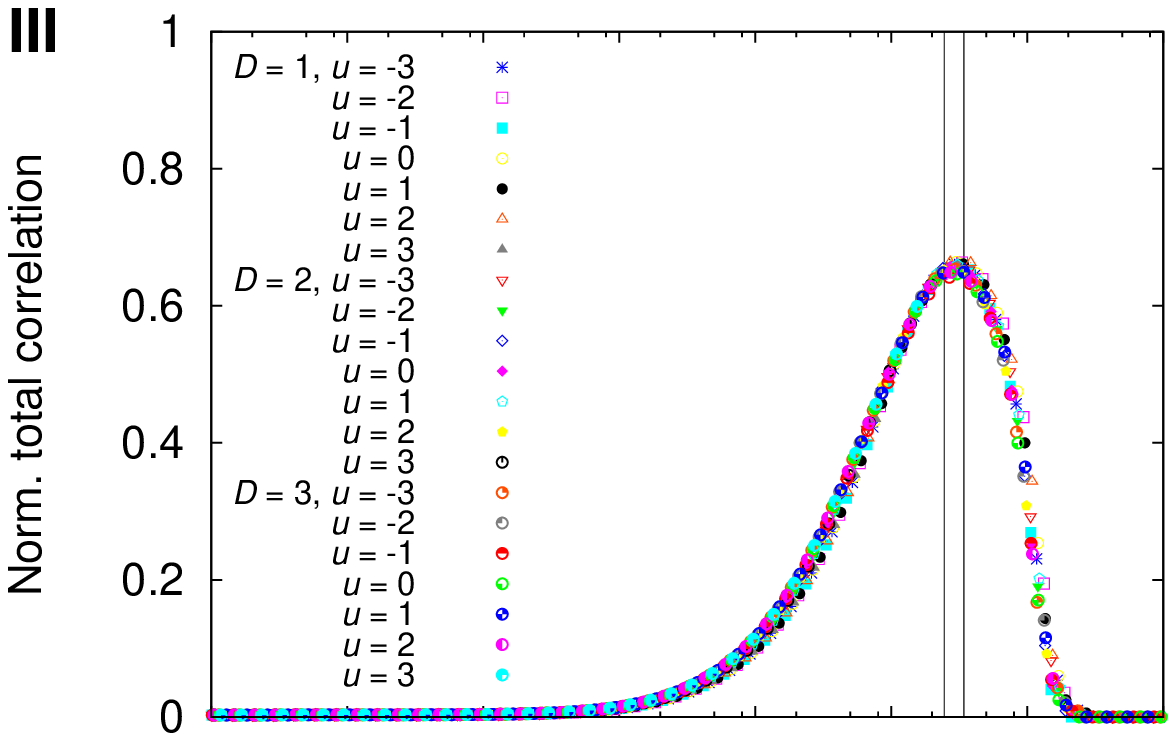}\\
\vspace{-20pt}
\includegraphics[scale=0.54]{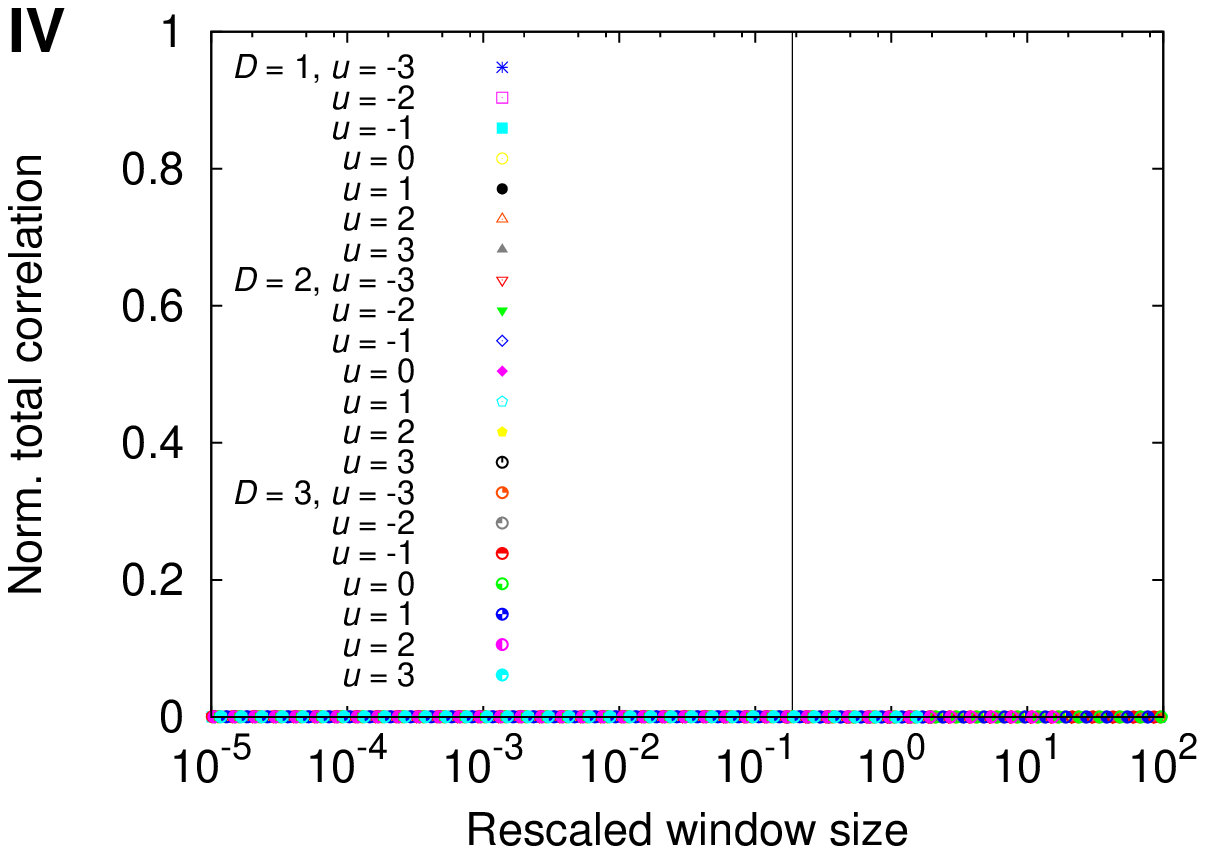}
\caption{Normalized total correlation ($C(\mathbf{X}^{(w)})/(N-1)$) against
rescaled window size ($w/T_0$) for the $I<0$ cases. I--IV identify settings. In
the keys to the plots, $u=\log_{10}\ell$. Vertical lines mark the smallest and
largest $w/T_0$ values given in Table~\ref{table2} (the smallest such value for
setting~IV is off range).}
\label{figure1}
\end{figure}

\begin{figure}
\centering
\includegraphics[scale=0.54]{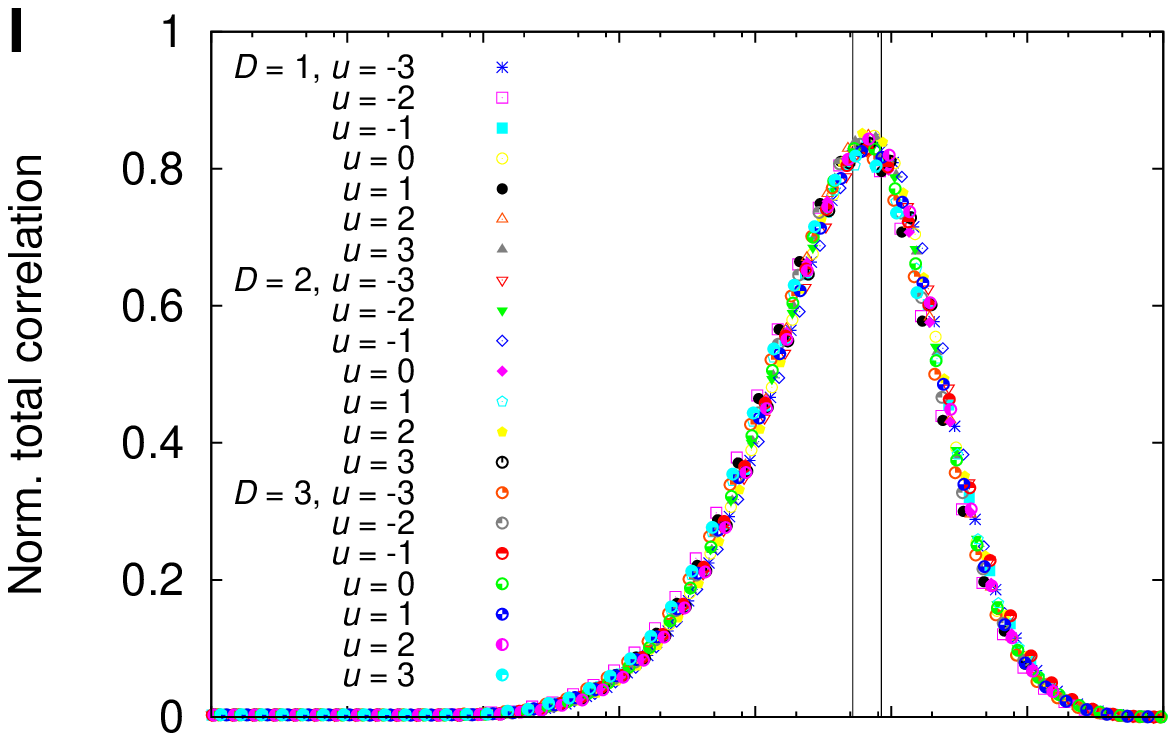}\\
\vspace{-20pt}
\includegraphics[scale=0.54]{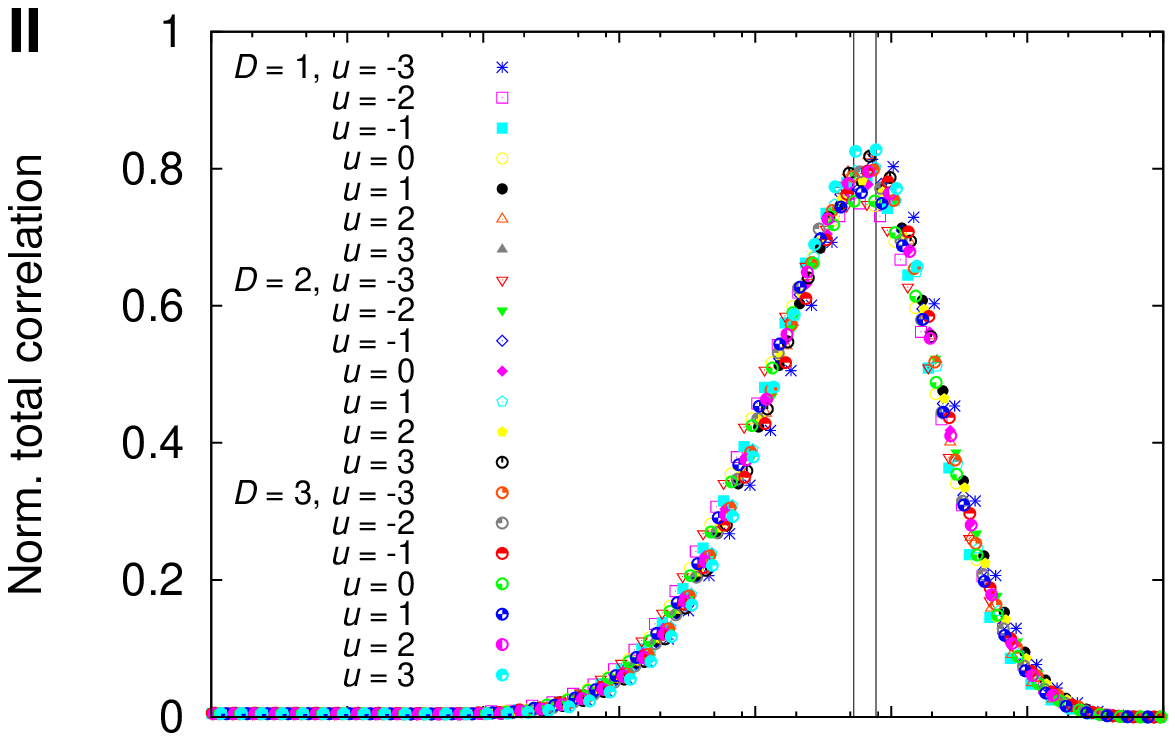}\\
\vspace{-20pt}
\includegraphics[scale=0.54]{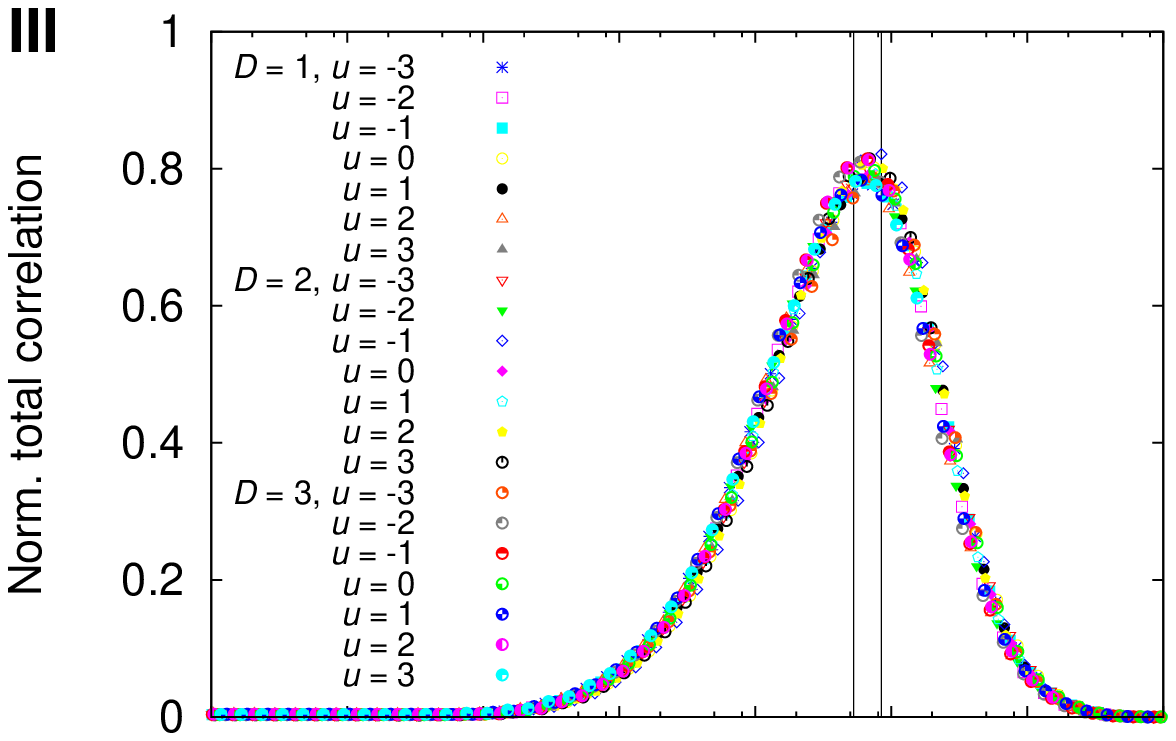}\\
\vspace{-20pt}
\includegraphics[scale=0.54]{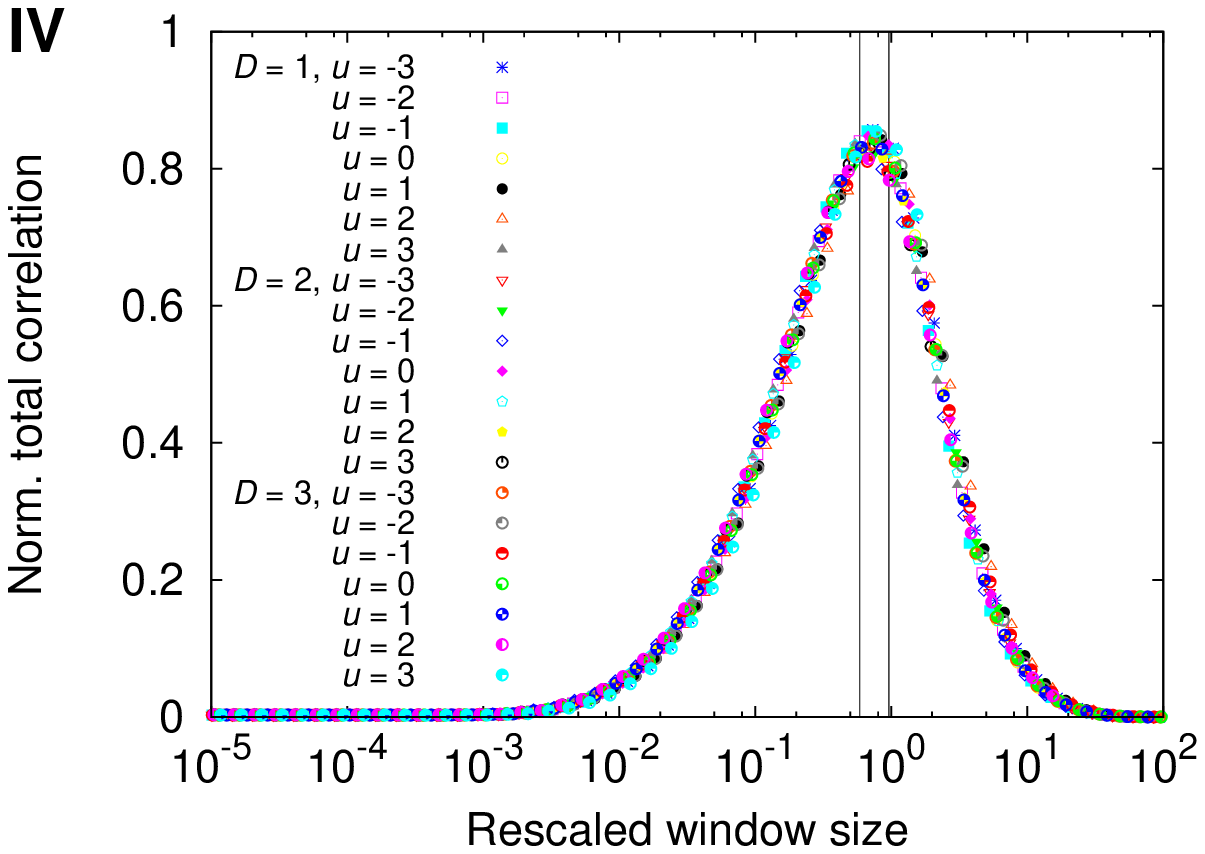}
\caption{Normalized total correlation ($C(\mathbf{X}^{(w)})/(N-1)$) against
rescaled window size ($w/T_0$) for the $I=0$ cases. I--IV identify settings. In
the keys to the plots, $u=\log_{10}\ell$. Vertical lines mark the smallest and
largest $w/T_0$ values given in Table~\ref{table2}.}
\label{figure2}
\end{figure}

\begin{figure}
\centering
\includegraphics[scale=0.54]{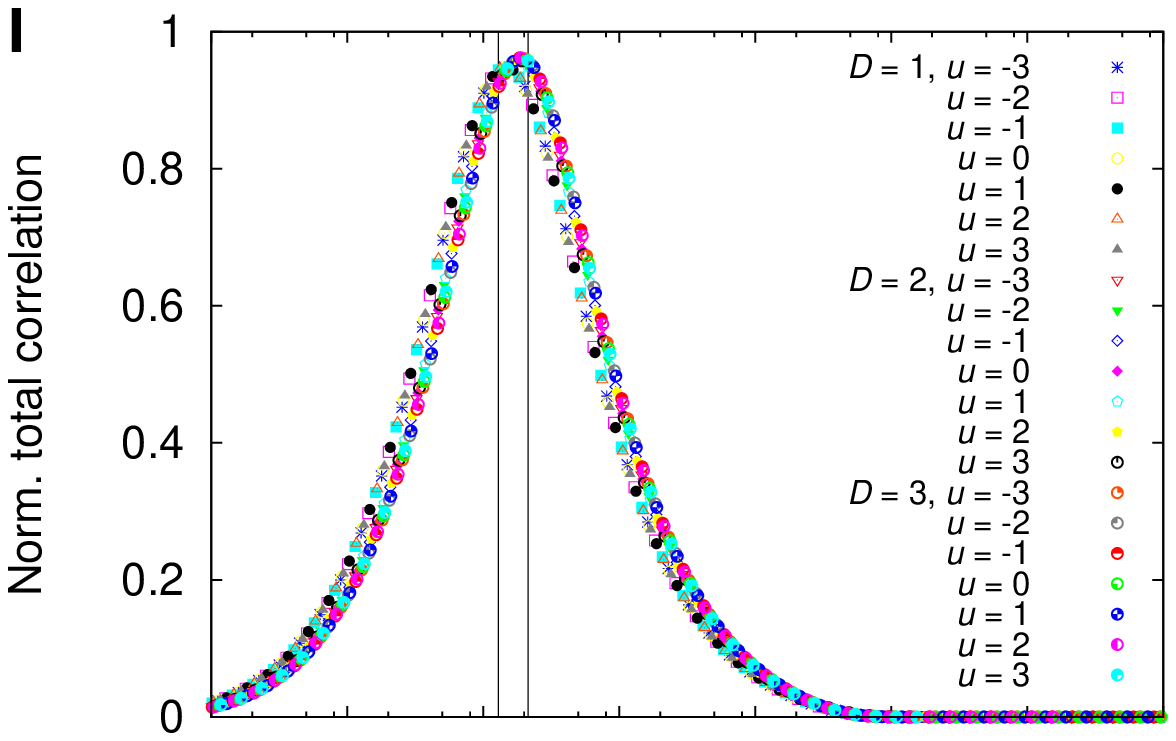}\\
\vspace{-20pt}
\includegraphics[scale=0.54]{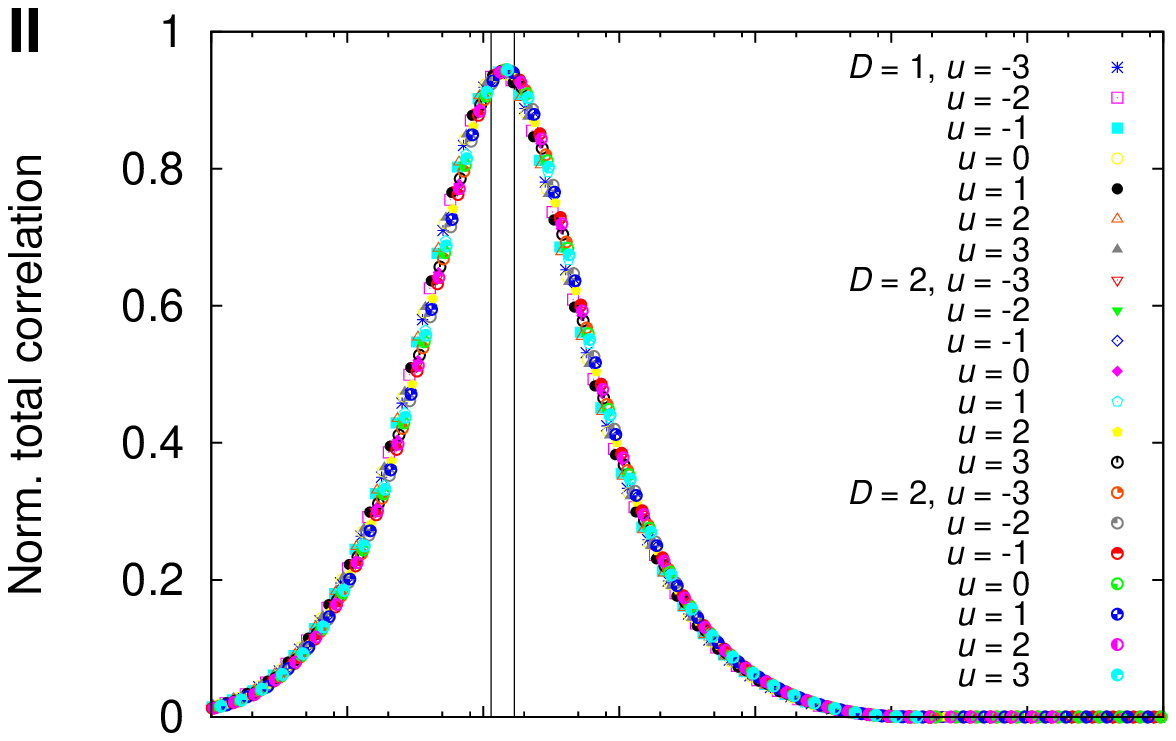}\\
\vspace{-20pt}
\includegraphics[scale=0.54]{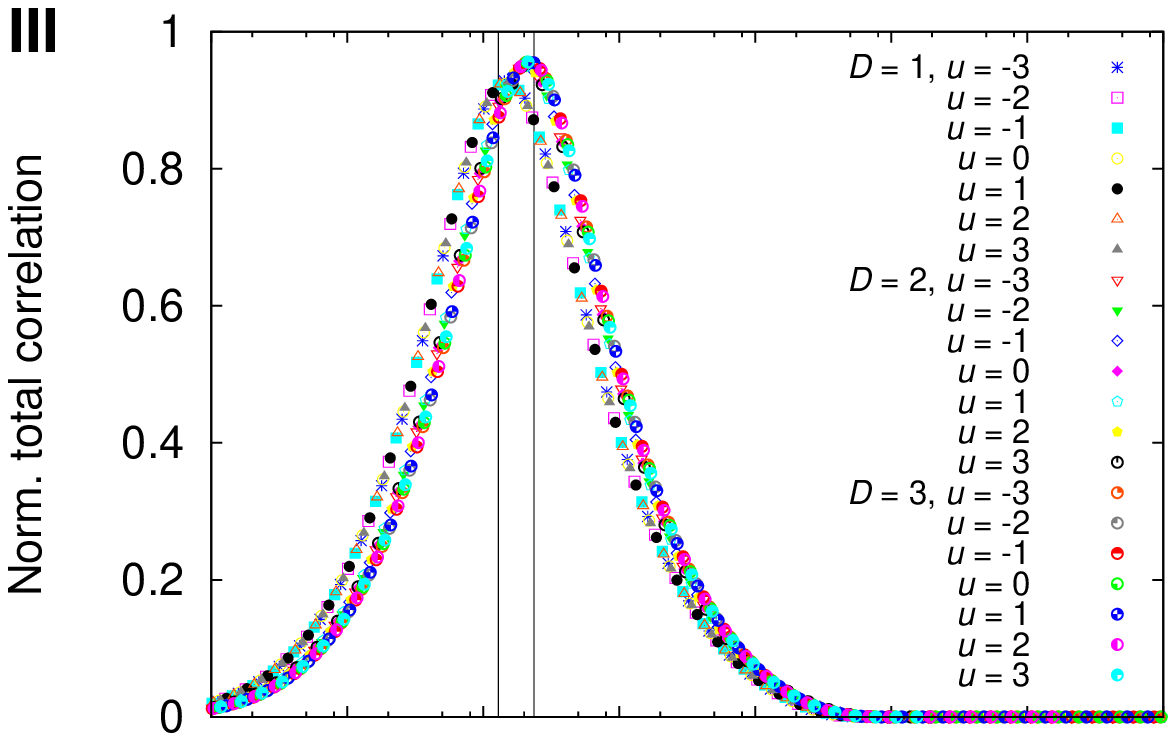}\\
\vspace{-20pt}
\includegraphics[scale=0.54]{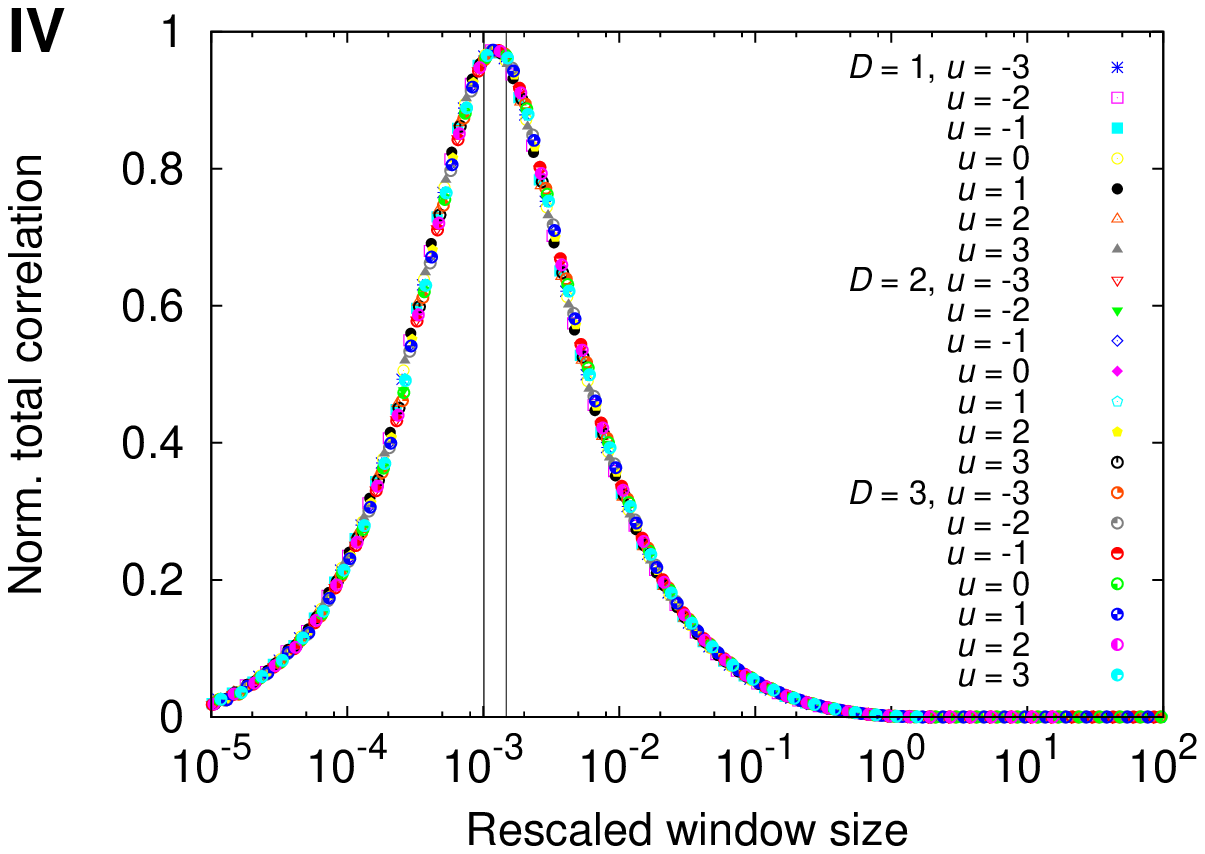}
\caption{Normalized total correlation ($C(\mathbf{X}^{(w)})/(N-1)$) against
rescaled window size ($w/T_0$) for the $I>0$ cases. I--IV identify settings. In
the keys to the plots, $u=\log_{10}\ell$. Vertical lines mark the smallest and
largest $w/T_0$ values given in Table~\ref{table2}.}
\label{figure3}
\end{figure}

The average value of the total correlation $C(\mathbf{X}^{(w)})$ resulting from
the computational experiments described in Section~\ref{exp} are presented in
normalized form (i.e., relative to the maximum $N-1$) in Figures~\ref{figure1},
\ref{figure2}, and~\ref{figure3}, respectively for the $I<0$ cases, the $I=0$
cases, and the $I>0$ cases. Each figure comprises four panels, numbered I--IV to
match the four settings of Table~\ref{table1}. Each panel contains $21$ plots,
each corresponding to one of the possible variations in the number of dimensions
$D$ and in the cube side $\ell$.

Each plot in the three figures is given against the rescaled version of the
window size $w$ given by $w/T_0$, where we recall from Eq.~(\ref{t0}) that $T_0$
is the delay a message incurs when traversing the distance $\Delta_\ell^{(D)}$
at speed $\sigma$. That is, the abscissas in
Figures~\ref{figure1}--\ref{figure3} are all relative to the geometric and
kinetic underpinnings to which each plot refers. This rescaling reveals that,
for each fixed combination of a setting with an imbalance profile (i.e., for
each panel in the three figures), the behavior of the average
$C(\mathbf{X}^{(w)})$ is essentially invariant with respect to how the various
$D$ and $\ell$ values are paired. Our choice in Section~\ref{imb} of $T_0/I$ as
the system's time constant is then clearly justified, as the value of $I$ is
fixed in each of the figures' panels. Moreover, such invariance also backs up
our choice of $\sigma=1$ for all simulations (see Section~\ref{exp}), since the
choice of any other value would simply alter the rescaling factors.

Nevertheless, there are signs in Figures~\ref{figure3}(I) and~(III) that such
invariance may not hold quite as well when $D=1$. We attribute this to the fact
that $\Delta_\ell^{(D)}$ is only an expected distance and as such affects the
role of $T_0$ as a rescaling factor differently for each number of dimensions
$D$. In particular, recalling from Section~\ref{geom} that the corresponding
variances get larger as $D$ is decreased, it seems clear that what is affecting
invariance in the two figures in question is precisely the poor
representativeness of $\Delta_\ell^{(1)}$. Even so, we see in
Figure~\ref{figure3} that this affects setting~III much more severely than it
does setting~I. The reason for this has to do with the value of imbalance $I$ in
each case, as we discuss in Section~\ref{disc}.

In all but the case of Figure~\ref{figure1}(IV), the average total correlation
starts off at some negligibly small value, then slowly climbs toward a peak
along an increase in $w/T_0$ by some orders of magnitude, and finally recedes as
$w/T_0$ is made to vary by further orders of magnitude. The plots in each of the
panels of Figures~\ref{figure1}--\ref{figure3} are given against a backdrop of
two vertical lines, the leftmost one marking the smallest $w/T_0$ at which total
correlation peaks for some $(D,\ell)$ pair, the rightmost marking the largest
such value. (The smallest $w/T_0$ value in the setting-III panel of
Figure~\ref{figure1} does not take into account the cases $D=2,3$ with
$\ell=10^3$, whose peaks seem to occur past $w=2^{10}$, the largest window used
in the simulations, and are therefore unknown.)

The backdrop lines' locations are detailed in Table~\ref{table2}, where the
$(D,\ell)$ giving rise to each one is also shown. The table also contains for
each such location a value for $\alpha$, the proportionality constant that
through Eq.~(\ref{aM0T0}) is used to relate the global message output $M_t(w)$
inside a size-$w$ window beginning at time $t$ to the reference output $M_0T_0$.
For $I=0$, we have found in Section~\ref{analysis} that the $w=w(\alpha)$
necessary for Eq.~(\ref{aM0T0}) to hold is time-invariant and given as in
Eq.~(\ref{aforI=0}), whence $\alpha=w/T_0$. This is reflected in
Table~\ref{table2}.

\begin{sidewaystable}
\caption{Smallest and largest $w/T_0$ at which $C(\mathbf{X}^{(w)})$ peaks for
each combination of a setting with a possibility for $I$ relative to $0$, all
configurations in each combination considered. Each of the two extremes occurs
for the configuration to which the $D$ and $\ell$ values shown refer. For each
of the two extremes, the value of $\alpha$ reported alongside it is such that
$w=\bar{w}(\alpha)$. The smallest $w/T_0$ reported for the $I<0$, setting-III
case does not take into account the peaks for $D=2,3$ with $\ell=10^3$, which
have remained undiscovered by the simulations.}
\label{table2}
\centering
\begin{tabular}{ccccccccccc}
\hline
Imbalance &Setting &\multicolumn{4}{c}{Smallest $w/T_0$ for $C(\mathbf{X}^{(w)})$ peak} &&\multicolumn{4}{c}{Largest $w/T_0$ for $C(\mathbf{X}^{(w)})$ peak}\\
\cline{3-6}\cline{8-11}
&&$D$ &$\ell$ &$w/T_0$ &$\alpha$ &&$D$ &$\ell$ &$w/T_0$ &$\alpha$\\
\hline
$I<0$ &I &3 &$10^{-3}$ &1.47582 &0.710 &&1 &$10^3$ &2.17223 &1.073\\
&II &3 &$10^{-2}$ &0.41743 &0.116 &&1 &10 &0.84853 &0.302\\
&III &2 &$10^2$ &2.45490 &1.395 &&3 &10 &3.41955 &1.917\\
&IV &2 &$10^3$ &$2\times 10^{-13}$ &0.050 &&2 &$10^{-2}$ &0.18729 &0.050\\
\hline
$I=0$ &I &3 &$10^{-3}$ &$=\alpha$ &0.522 &&2 &$10^{-1}$ &$=\alpha$ &0.848\\
&II &1 &1 &$=\alpha$ &0.530 &&3 &$10^3$ &$=\alpha$ &0.774\\
&III &2 &$10^{-2}$ &$=\alpha$ &0.530 &&2 &$10^{-1}$ &$=\alpha$ &0.848\\
&IV &1 &$10^{-2}$ &$=\alpha$ &0.586 &&1 &$10^2$ &$=\alpha$ &0.960\\
\hline
$I>0$ &I &1 &$10^{-1}$ &0.00129 &0.009 &&3 &$10^3$ &0.00214 &0.015\\
&II &1 &$10^{-2}$ &0.00114 &0.008 &&2 &$10^2$ &0.00170 &0.012\\
&III &1 &$10^{-1}$ &0.00129 &0.009 &&3 &10 &0.00236 &0.017\\
&IV &1 &$10^{-3}$ &0.00101 &0.007 &&3 &1 &0.00148 &0.010\\
\hline
\end{tabular}

\end{sidewaystable}

For $I\neq 0$, on the other hand, satisfying Eq.~(\ref{aM0T0}) for some fixed
$\alpha$ requires $w$ to vary with time according to Eq.~(\ref{wt}). This is
what one would expect, since the dwindling message output that results from an
$I<0$ scenario requires an ever larger window size to accommodate the amount of
traffic given by $\alpha M_0T_0$, and likewise the output expansion caused by
$I>0$ requires progressively smaller window sizes. This dependency on time is
reflected in the inequalities of Eqs.~(\ref{wforIneg}) and~(\ref{wforIpos}),
respectively for $I<0$ and $I>0$, where the earliest window size ($w_0$), as
well as the average one ($\bar{w}$) and the latest ($w_*$) are put in
perspective. The values of $\alpha$ given in Table~\ref{table2} for the
$I\neq 0$ cases come from Eq.~(\ref{barw}) by letting each value of $w/T_0$
reported in the table be such that $w=\bar{w}(\alpha)$.

\begin{figure}
\centering
\includegraphics[scale=0.54]{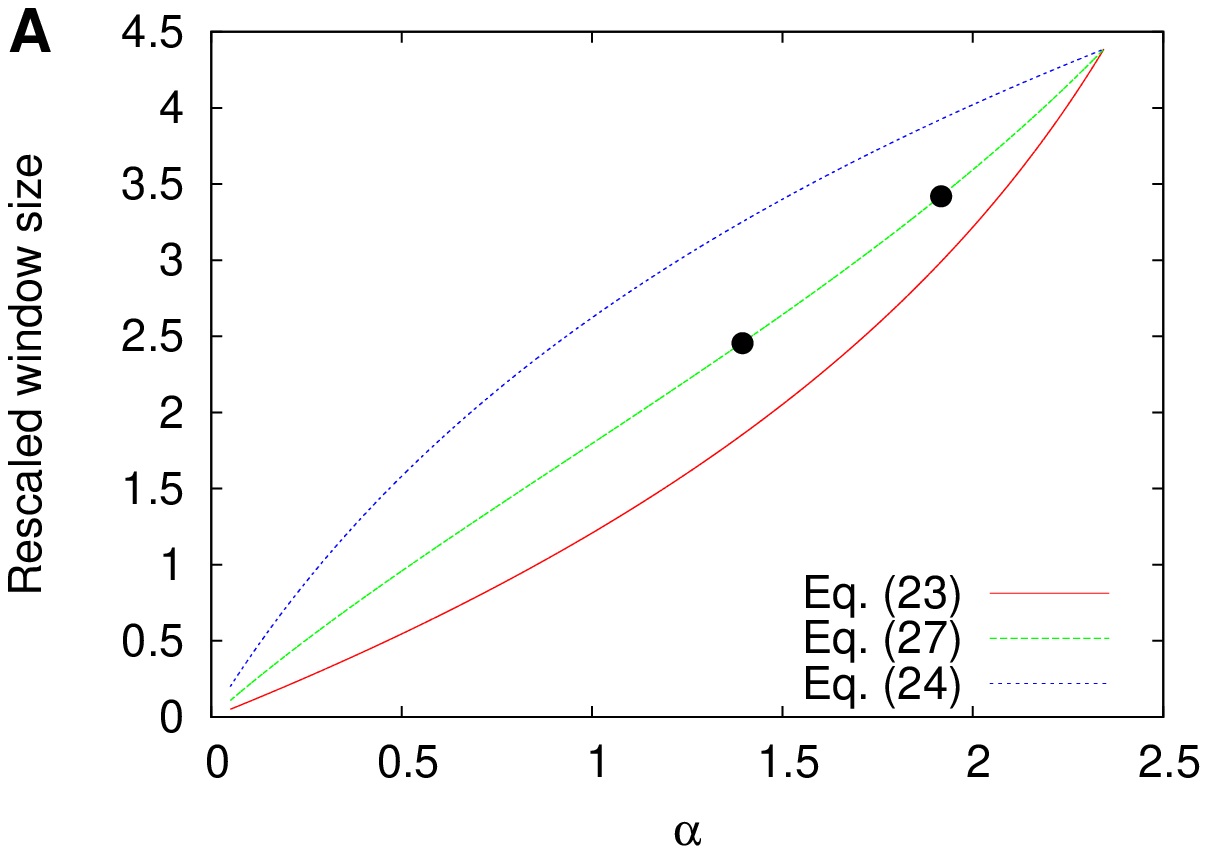}\\
\includegraphics[scale=0.54]{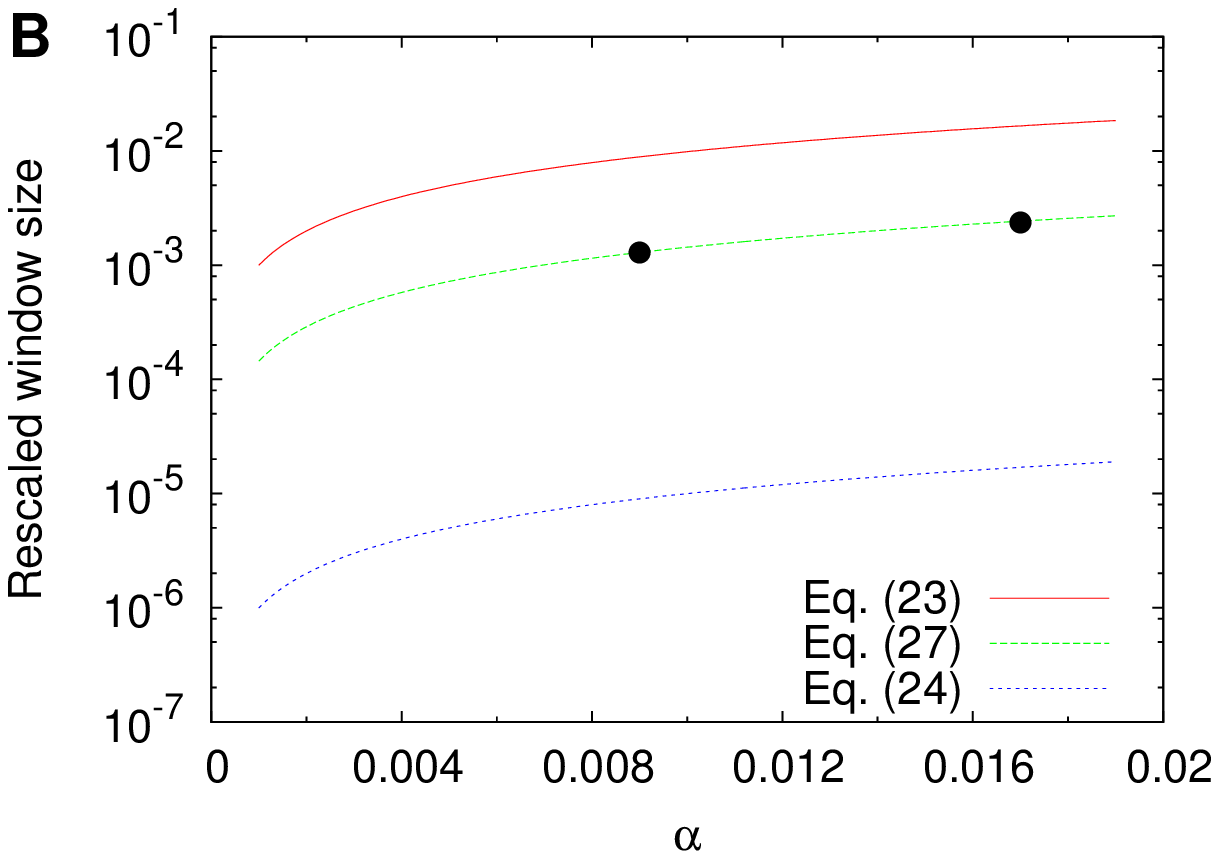}
\caption{Rescaled window size ($w/T_0$) against $\alpha$, as per
Eqs.~(\ref{w0}),~(\ref{barw}), and~(\ref{w*}), respectively for $w=w_0$,
$w=\bar{w}$, and $w=w_*$. All plots refer to setting~III, with either $I<0$ (A)
or $I>0$ (B). The corresponding $(\alpha,w/T_0)$ pairs in Table~\ref{table2} are
highlighted as bullets. Plot ranges are given by the intervals $[0.05,M/M_0)$ in
A (to within a difference of $10^{-4}$ on the upper side) and $[0.001,0.019]$ in
B.}
\label{figure4}
\end{figure}

This use of Eq.~(\ref{barw}) requires the expected value of $M$ (the number of
messages sent in a run) to be estimated from the $100$ independent trials for
each configuration of interest. We do this by first averaging the number of
messages received during each run (let $E$ denote this average) and then
recalling that each message received entails an expected number of messages sent
equal to $(N-1)p_\to/\mu$. Our estimate of the expected $M$ is
then\footnote{Our simulator is event-based, each event being the reception of a
message. The total we have available at the end of each simulation is that of
messages received, not sent, and the two may differ because not every message
sent gets received, by virtue of the preestablished maximum number of messages
in transit having been reached (see Section~\ref{methods}). We then need this
workaround to find the expected $M$ through $E$.}
\begin{equation}
M=M_0+\frac{E(N-1)p_\to}{\mu}.
\end{equation}
An illustration highlighting the use of Eq.~(\ref{barw}) on the setting-III
cases of Table~\ref{table2} for $I\neq 0$ is given in Figure~\ref{figure4},
where the plots relative to Eq.~(\ref{w*}) rely on the same estimate of the
expected $M$ as above.

\section{Discussion}
\label{disc}

Even though our notation for total correlation, $C(\mathbf{X}^{(w)})$, may
suggest that it is a function solely of the $N$ random variables in
$\mathbf{X}^{(w)}$, it is in fact a function of the joint probability mass
function associated with those variables as well. Therefore, changing the
allotment of probability mass to the various points in $\{0,1\}^N$ must have an
impact on the value of $C(\mathbf{X}^{(w)})$ as a matter of principle. Thus,
while we know that such variation must never lead total correlation to fall
below zero or above $N-1$, figuring out the resulting expected value has the
potential of aiding in the interpretation of any particular figure one
encounters in some situation of interest. As far as we know, such expected value
has so far remained unknown, but resorting to results on the expected value of
the joint Shannon entropy ($H(\mathbf{X}^{(w)})$, in our case) \cite{c95}, it is
possible to ascertain that an upper bound on the expected value of
$C(\mathbf{X}^{(w)})$ is approximately $0.6$ \cite{nb11}.

Once normalized to $N-1$, this upper bound would be unnoticeable in any of the
panels in Figures~\ref{figure1}--\ref{figure3}. In fact, it would fall two
orders of magnitude below the nearly flat values of Figure~\ref{figure1}(IV),
which are of the order of $10^{-1}$. Clearly, then, in all of settings~I--IV and
for all three imbalance scenarios, our model is seen to be promoting the rise of
probability mass functions leading to total correlations significantly above the
expected value. In most cases, this occurs over ranges for the value of $w$,
the temporal-window duration that provides meaning to the random variables in
$\mathbf{X}^{(w)}$, spanning several orders of magnitude.

Not only this, but in most cases we have found normalized total correlation to
peak at substantial levels: between $0.22$ and $0.67$ in settings~I and~III when
$I<0$ (Figure~\ref{figure1}); $0.77$ and $0.86$ in all four settings when $I=0$
(Figure~\ref{figure2}); and $0.93$ and $0.97$ in all four settings when $I>0$
(Figure~\ref{figure3}). The peaks relating to the $I=0$ cases (and also those of
the $I>0$ cases, but to a more limited extent) seem to occur largely
independently of the setting in question, that is, regardless of the number of
nodes $N$ or the parameters that control firing (the probability $p_-$ and the
threshold $\tau$).

Understanding this independence comes from focusing on the $I=0$ cases, since
for $I=0$ the value of probability $p_\to$ varies from setting to setting as a
function of $N$ and $\mu$ (therefore a function of $p_-$ and $\tau$ as well, by
Eq.~(\ref{mu})); see Table~\ref{table1}. Setting $p_\to$ in this way aims
precisely to cause $I=0$ and, as a consequence, tends to compensate for any
dependency of total-correlation peaks on $N$, $p_-$, or $\tau$. It also suggests
that the value of $I$, along with that of $w/T_0$, is a major player when it
comes to determining how much total correlation can be achieved. However,
increasing the value of $p_\to$ from those used to ensure $I=0$ to $p_\to=0.06$,
thus ensuring $I>0$, preserves nearly the same independence of peak values as
when $I=0$ but fails to keep $I$ constant from setting to setting (see
Table~\ref{table3} for the specific value of $I$ in each one). With the
increased value of $p_\to$ and the ever-growing barrage of messages that ensues,
this independence is now supported by window sizes at least two orders of
magnitude lower (though yielding taller peaks) but not significantly by the
value of $I$ itself. The value of $I$, however, does make its influence felt, in
the following manner. As mentioned in Section~\ref{results}, rescaling window
size $w$ through a division by $T_0$ makes some of the $I>0$ cases stand out for
$D=1$ by failing to comply (though to a small degree) with the invariance to the
values of $D$ and $\ell$ that seems to be the rule. In that section we
correctly attributed this to the increased variance of $T_0$ (through that of
$\Delta_\ell^{(1)}$), but the deviation from invariance clearly increases with
the value of $I$ as well ($1.81$ and $3.06$, by Table~\ref{table3}, respectively
for settings~I and~III).

\begin{table}
\caption{Nonzero imbalance values for each of settings~I--IV.}
\label{table3}
\centering
\begin{tabular}{ccc}
\hline
Setting  &$I$ for $p_\to=0.01$  &$I$ for $p_\to=0.06$\\
\hline
I  &$-0.53$  &1.81\\
II  &$-0.72$  &0.68\\
III  &$-0.32$  &3.06\\
IV  &$-0.78$  &0.29\\
\hline
\end{tabular}

\end{table}

On the other hand, the cases for which $I<0$ are different in regard to this
independence issue. In fact, it is clear from Figure~\ref{figure1} that
total-correlation peaks are highly affected by the setting in question. In
particular, they are increasingly lower than the corresponding peaks for $I=0$,
and also occur for increasingly lower values of $w/T_0$, as the value of $I$
(see Table~\ref{table3}) becomes more negative. For fixed $T_0$, therefore, this
consolidates the value of $I$ as the main driver defining the window duration
for which total correlation peaks, provided $I\le 0$.

The $I=0$ cases are particularly interesting also because total correlation was
in all configurations of all settings found to peak roughly between
$w/T_0=0.522$ and $w/T_0=0.96$ (see Table~\ref{table2}). As we noted in
Section~\ref{results}, this is also the range of $\alpha$, which for $I=0$ is
the fraction of the reference message output $M_0T_0$ yielding the message
traffic inside a window of duration $w$ beginning, essentially, at any time
$t\ge 0$. That total correlation should peak with $\alpha$ between roughly
$0.522$ and $0.96$ seems well aligned with our discussion at the end of
Section~\ref{tc} on maximizing total correlation, through which we found out
that a condition necessary (though not sufficient) to such maximization is that
only half the nodes receive messages inside the window.

Relativizing message traffic inside the window to $M_0T_0$ in the $I\neq 0$
cases must be approached differently, because now fixing $\alpha$ requires the
duration of the window starting at time $t$ to either expand with $t$ (if $I<0$)
or shrink with it (if $I>0$). In these cases, taking each $w$ in
Table~\ref{table2} to be the average duration of these windows reveals that
total correlation peaks for $w/T_0$ between about $1.47582$ and $3.41955$ for
$I<0$ (therefore somewhat above the range recorded for the $I=0$ cases),
ignoring the more or less degenerate cases of settings~II and~IV, and between
about $0.00101$ and $0.00236$ for $I>0$ (therefore substantially below the $I=0$
ranges, as noted above). The corresponding values of $\alpha$, given by
Eq.~(\ref{barw}) and illustrated in Figure~\ref{figure4} for setting~III, range
between $0.71$ and $1.917$ for $I<0$, between $0.007$ and $0.017$ for $I>0$.

\subsection{What of the brain?}
\label{brain}

Right at the opening of Section~\ref{intro} we mentioned the brain as the most
representative threshold-based system we know. It is only fitting, then, that we
should try and relate it more closely to the model we have developed and
analyzed in this paper. This is no simple task for at least two reasons. The
first has to do with the model itself, which as we also mentioned in
Section~\ref{intro} has not been meant to faithfully represent the specifics of
any one threshold-based system. The second reason is that very little has so far
been found out of the brain's so-called microconnectome \cite{bk12}, that is,
its network structure at the cellular level, which is where thresholds work
their influence.

Here we resort to the few glimpses that are available in order to estimate
parameter values within our model as well as possible. The largest
microconnectome to have been mapped to date is from the V1 area of the mouse
visual cortex \cite{lbrghgr16}. It reveals a total of $1\,278$ excitatory
neurons and $581$ inhibitory neurons, which following our discussion in
Section~\ref{rgraph} allows as to conclude
$p_-\approx 581/(581+1\,278)\approx 0.31$, thus agreeing with the commonly
accepted range of values for the ratio of inhibitory neurons \cite{rflh10}. It
also reveals a total of $29$ connected pairs involving $45$ neurons, which again
in the spirit of Section~\ref{rgraph} leads to
$p_\to\approx 29/(45\times 44)\approx 0.015$. Using $N=45$ and $\tau=7.5$
(halfway between $5$ and $10$, which seem to delimit the range of the typical
neuron's ratio of threshold potential to synaptic potential
\cite{ stg01,hshkdm05,shdym06}), we obtain an imbalance $I=-0.96$, close
therefore to the minimum possible imbalance ($I=-1$).
 
Note that concluding $p_\to\approx 0.015$ reflects an effective, rather than
merely anatomical, view of the microconnectome. That is, the $29$
axon-to-dendrite connections leading to this estimate of the value of $p_\to$ do
not result simply from the possibility that action potentials travel from a
neuron's axon toward one of another's dendrites, but rather from actually
observing such potentials.\footnote{To further strengthen this notion that the
value of $p_\to$ is highly network-dependent in terms of effective signaling in
the brain, we note that a completely different estimate of $p_\to$ can be
deduced from a slightly earlier (but similar in the sense of targeting effective
connections) study \cite{sb15}, whose authors report a neuron's expected
out-degree to be in the order of $10^0$. The study in question is based on the
rodent somatosensory cortex, but should a similar conclusion hold for the human
cortex taken as a whole, with its $16$ billion neurons \cite{hh09}, then an
estimate of $p_\to$ would place its value at the order of $10^{-10}$ and that of
$I$ very close indeed to $-1$. This is all rather speculative, though,
especially if we consider the already established fact that different cortical
areas of the primate brain have different neuronal densities \cite{caylk10}.}
This distinction is of paramount importance in the present study, since the
parameter $p_\to$ is semantically tied to the actual sending of messages in our
model (once interpreted as a directed graph, as in Section~\ref{rgraph}), not
merely to the existence of a channel through which such messages could be sent.

To judge from the values of $I$ for $p_\to=0.01$ in Table~\ref{table3}, and from
the conclusions we have drawn regarding the $I<0$ cases of settings~I--IV, this
entire attempt to interpret cortical activity in the light of our model seems to
be suggesting that we view it as a threshold-based system severely imbalanced
on the negative side, therefore incapable of giving rise to any significant
amount of total correlation. The missing component, of course, is that the
cortex is continually subjected to new input, either external or originating
spontaneously from within \cite{lbh13}, hence a better characterization of
local imbalance in this case would be based on
\begin{equation}
I'=\frac{(N-1)p_\to+\mu_0}{\mu}-1,
\end{equation}
with $\mu_0$ standing for some core amount of external or spontaneous input
expected to accompany every $\mu$ messages incoming from other cortical neurons.
Clearly, $\mu_0>0$ implies $I'>I$ whenever all else remains constant. Local
imbalance would then be defined through $I'\neq 0$, but without the availability
of some rationale on which to base an estimate of $\mu_0$ this is essentially as
far as we can go. In any event, such redefined imbalance might lie above the
$I<0$ values of Table~\ref{table3} while still being negative.\footnote{Letting
$I^*$ denote the greatest of these negative values for $I$, the purported $I'$
such that $I^*<I'<0$ would require $0<\mu_0<\mu-(N-1)p_\to$, assuming for the
former of these inequalities that $N$ and all parameters remain unchanged
between $I^*$ and $I'$.} In this case, we could expect $C(\mathbf{X}^{(w)})$ to
peak for $w/T_0$ a few orders of magnitude above $3.41955$ (see
Table~\ref{table2}).

We can go further and estimate the value of $T_0$ as well. Because the cerebral
cortex takes up $82\%$ of all brain mass in humans \cite{hh09}, and assuming a
uniform density in addition to a brain volume of
$1.5\times 10^{-3}\ \textrm{m}^3$ \cite[BNID 112053]{mjmws10}, we can take the
cortical volume to be roughly $1.23\times 10^{-3}\ \textrm{m}^3$. In our model,
this volume corresponds to the three-dimensional cube of side
$\ell\approx 0.107\ \textrm{m}$, whence
$\Delta_\ell^{(3)}\approx 0.071\ \textrm{m}$ (see Section~\ref{geom}). Assuming
further that action potentials propagate on myelinated axons at a speed
($\sigma$) between $50$ and $100\ \textrm{m/s}$ \cite[BNID 107125]{mjmws10}
\cite{sw12} leads, by Eq.~(\ref{t0}), to a value of $T_0$ roughly between $0.71$
and $1.42\ \textrm{ms}$. This would place $w$, the time window duration for peak
total correlation, a few orders of magnitude above some value between
$2.4$ and $4.9\ \textrm{ms}$. Significant total correlation also arises for
$w/T_0$ values below or above the peak's location by a small factor, so window
durations of a few hundred milliseconds would support information integration as
defined. These, as it turns out, are time lapses within range of the commonly
accepted delay before a percept can be rendered conscious \cite{hks16}. 

\section{Conclusion}
\label{concl}

Our approach to quantifying the integration of information to which interacting,
distributed threshold-based units give rise has relied on a very general model,
tailored to no specific system in particular but built around three fundamental
notions: that the units interact with one another by passing positively or
negatively tagged messages among them; that a unit sends messages out as a
function of how the tag balance of incoming messages relates to a threshold; and
that this sending out of messages is selective, in the sense of being
addressable to a subset of the units only. Assuming that the units are
positioned in a one-, two-, or three-dimensional cube uniformly at random, and
moreover considering a temporal window of duration $w$, we have demonstrated by
means of extensive computational experiments that information does get
integrated in significant amounts inside the window, depending chiefly on the
value of $w$, on the average delay incurred by a message, and on local message
imbalance (how much gets transmitted vis-\`a-vis what is received, itself
dependent on the combined effect of the message-passing and threshold
parameters). We have analyzed situations of peak information integration and
related the results to cortical dynamics by fixing the model's parameters
accordingly. This has served to suggest a validation of the model, despite its
purposeful generality, highlighting its potential usefulness for systemic
studies of threshold-based interacting units.

Given the window duration $w$ and the set of random variables $\mathbf{X}^{(w)}$
(one variable per unit, each related to the reception of messages by the unit in
question within the window), our measure of information integration has been
the total correlation of the variables in $\mathbf{X}^{(w)}$, here denoted by
$C(\mathbf{X}^{(w)})$. This measure seeks to quantify the interdependence of the
variables on one another and can be interpreted as information gain that is
attained in excess of the total gain the units achieve at the local level. Total
correlation can also be interpreted as the Kullback-Leibler (KL) divergence of
one probability mass function relative to another, viz., of
$\mathrm{Pr}(\mathbf{X}^{(w)}=\mathbf{x})$ relative to
$\prod_{i=1}^N\mathrm{Pr}(X_i^{(w)}=x_i)$. That is, $C(\mathbf{X}^{(w)})$ can be
rewritten as
\begin{equation}
C(\mathbf{X}^{(w)})=
\sum_{\mathbf{x}\in\{0,1\}^N}
\mathrm{Pr}(\mathbf{X}^{(w)}=\mathbf{x})
\log_2\frac
{\mathrm{Pr}(\mathbf{X}^{(w)}=\mathbf{x})}
{\prod_{i=1}^N\mathrm{Pr}(X_i^{(w)}=x_i)}.
\end{equation}
This expression highlights the well-known fact that the KL divergence is
asymmetric with respect to the two mass functions it applies to. Note, however,
that this is of no import in our context, because total correlation corresponds
to the very specific case in which the KL divergence applies to the two mass
functions above and in the direction indicated only. We mention this issue
because the current version of the all-partitions theory of integrated
information mentioned in Section~\ref{intro} abandons the KL divergence in favor
of the so-called earth-mover's distance because of the latter's symmetry and
consequent status as a distance between two probability mass functions
\cite{oat14}. This provides further distinction between the two approaches. 

We finalize by noting that extensions to our approach are certainly possible,
especially by attempting to encompass those systems that, despite embodying
explicit references to a threshold-based dynamics, do so in a manner that is not
completely aligned with the one our model assumes. One example involves the role
of thresholds in the grouping and synchronization of the interacting units
\cite{kks15}. Perhaps our approach can provide useful insight in such contexts
as well.

\subsection*{Acknowledgments}

The author acknowledges partial support from CNPq, CAPES, and a FAPERJ BBP
grant.

\bibliography{infoint}
\bibliographystyle{unsrt}

\end{document}